\documentclass[
reprint,
 amsmath,amssymb,
 aps,
]{revtex4-1}

\usepackage{graphicx}
\usepackage{dcolumn}
\usepackage{bm}
\usepackage{hyperref}
\usepackage[mathscr]{euscript}
\usepackage{setspace}

\begin{document}

\preprint{APS/123-QED}

\title{Self-organization in soliton modelocked parametric frequency combs}

\author{Y. Henry Wen}
\email{yhw2@cornell.edu}

\author{Michael R. E. Lamont}

\author{Alexander L. Gaeta}
 \altaffiliation[Also at ]{The Kavli Institute at Cornell for Nanoscale Science, Cornell University, Ithaca, NY 14853, USA}
 \affiliation{School of Applied $\&$ Engineering Physics, Cornell University, Ithaca, NY 14853, USA}
 
 \author{Isabel M. Kloumann}
 \affiliation{Center for Applied Mathematics, Cornell University, Ithaca, NY 14853, USA}
\author{Steven H. Strogatz}
\affiliation{Center for Applied Mathematics, Cornell University, Ithaca, NY 14853, USA}

\date{\today}
\begin{abstract}
We show that self-organization occurs in the phase dynamics of soliton modelocking in parametric frequency combs. Reduction of  the Lugiato-Lefever equation (LLE) to a simpler set of phase equations reveals that this self-organization arises via mechanisms akin to those in the Kuramoto model for synchronization of coupled oscillators. In addition, our simulations show that the phase equations evolve to a broadband phase-locked state, analogous to the soliton formation process in the LLE. Our simplified equations intuitively explain the origin of the pump phase offset in soliton-modelocked parametric frequency combs. They also predict that the phase of the intracavity field undergoes an anti-symmetrization that precedes phase synchronization, and they clarify the role of chaotic states in soliton formation in parametric combs. 
\pacs{42.82.Et, 03.65.Xp, 42.65.Pc}

\begin{description}
\item[PACS numbers] 42.82.Et, 03.65.Xp, 42.65.Pc
\end{description}
\end{abstract}

\maketitle

\begin{figure}[t]
\centerline{\includegraphics[width=8cm]{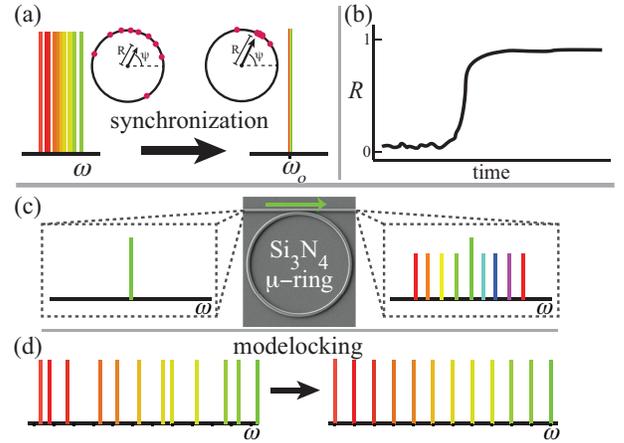}}
\caption{\label{fig:one} (a) Synchronization of a population of oscillators with non-equal natural frequencies to a phase-locked state with an identical frequency.  A large fraction of the oscillator phases lock to $\psi$, the average of all phases. (b)  Abrupt transition of the coherence $R$ to synchronization. (c) CW-pumped FWM generation of broadband frequency comb in silicon nitride micro-rings. (d) Modelocking of cavity modes results in equidistant frequency spacings between adjacent modes.}
\end{figure}
\begin{figure}[t]
\centerline{\includegraphics[width=8.5cm]{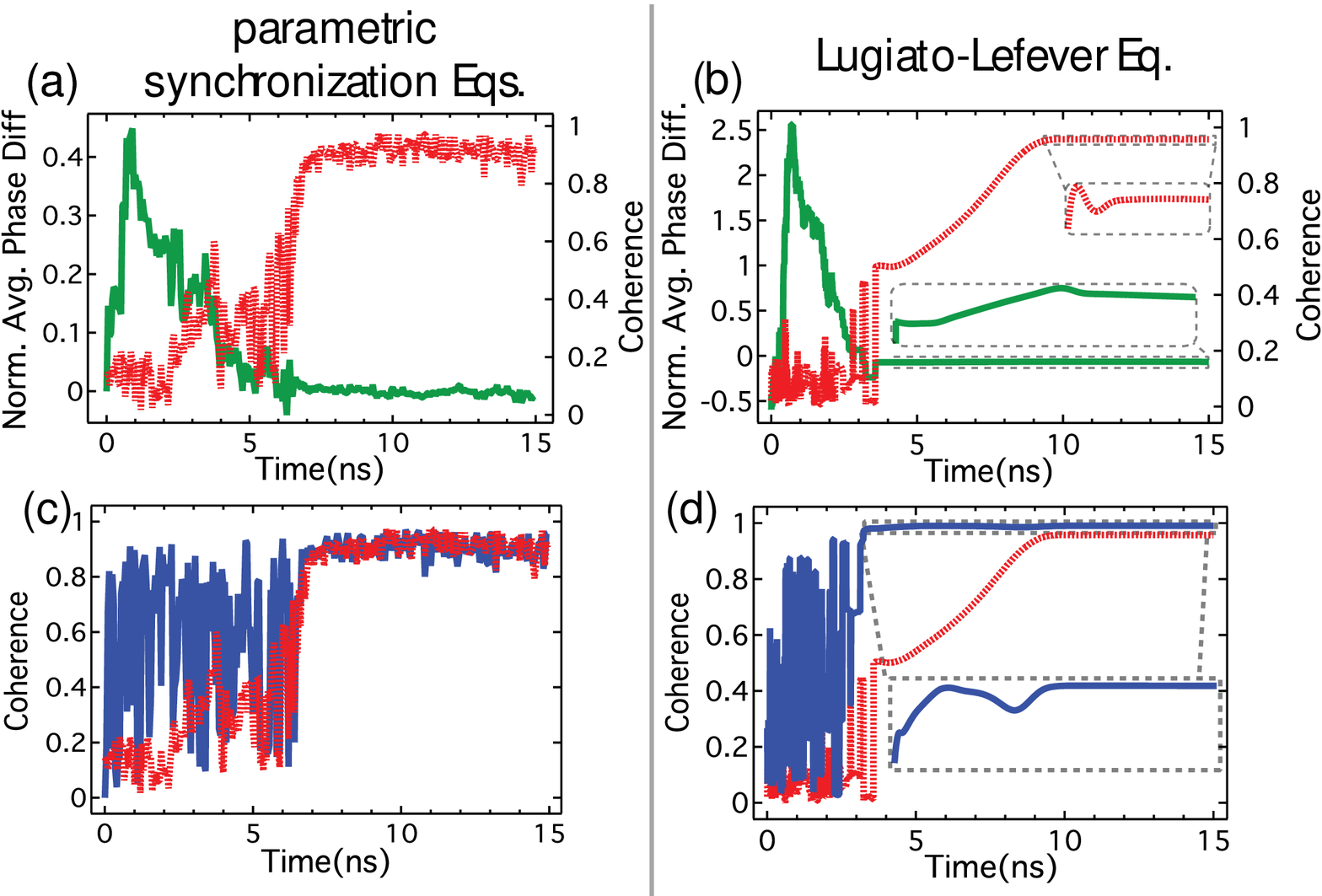}}
\caption{\label{fig:four} (a, b) Coherence $R$ (red-dotted) and normalized average phase difference $\theta_{o}$ (green). An abrupt transition to a stable phase-synchronized state is observed in both systems. (c, d) Coherence $R$ (red-dotted) and phase symmetry $R_{sym}$ (blue).} 
\end{figure}

\indent  A large collection of coupled oscillators with slightly different natural frequencies can undergo a transition to a phase-locked state with identical frequencies.  This phenomenon appears in many systems spanning biology, chemistry, neuroscience, and physics \cite{Kuramoto, Strogatz}. Examples include power grid networks, neural networks, chemical oscillators, and arrays of Josephson junctions and semiconductor lasers \cite{Motter, Wolmelsdorf, Kiss, Weisenfeld, Kozyreff}.  Self-organization in such systems has been modeled by the Kuramoto model, which describes the time-evolution of the phase $\phi_p(t)$ of an oscillator $p$ as an interaction between its natural frequency $\omega_p$ and its coupling to the phases of all the other oscillators. The governing equations are $\dot{\phi}_{p}= \omega_p+\kappa\sum^{N}_{m}\sin(\phi_m-\phi_p)$, where $\kappa$ is the coupling strength \cite{Strogatz}. This model can be recast in an order-parameter formulation, where an average phase $\psi$ and a coherence $R(t)$ are defined via $R(t)e^{i\psi}= \frac{1}{N}\sum^{N}_{m}e^{i\phi_m}$  (Fig. \ref{fig:one}a,b). Then the Kuramoto model becomes $\dot{\phi}_{p}= \omega_p+\kappa R(t)\sin(\psi-\phi_p)$. Viewed this way, $\phi_p$ is no longer coupled to every individual oscillator's phase, but only to the average phase $\psi$. Moreover the effective strength of the coupling is proportional to the coherence $R$. This proportionality between coupling and coherence creates a positive feedback which, for a sufficiently large $\kappa$, gives rise to an abrupt transition in which a macroscopic fraction of the oscillators' frequencies spontaneously synchronize. \\
\indent In optics an alternative form of phase locking can occur in lasers and parametric oscillators between a large collection of cavity modes with nearly equidistant frequency separations. In these systems the nearest-neighbor spacing across the modes varies due to dispersion within the cavity.  In the presence of nonlinearity within the cavity, the system can spontaneously modelock such that the frequency spacings between the oscillating modes become identical (Fig. \ref{fig:one}c,d).  This behavior has been studied extensively in the context of ultrashort pulse generation via soliton formation in lasers \cite{Haus, Kartner} and in micro-resonator-based four-wave mixing (FWM) parametric oscillators \cite{Saha, Herr2}. Such oscillators have attracted much recent interest since they can potentially be used for a wide range of applications in optical information storage and processing \cite{Leo}, broad-band frequency combs \cite{Yoshi}, optical spectroscopy, and frequency metrology \cite{Udem}. While several theoretical \cite{Coillet, Loh} and experimental studies \cite{DelHayeHot,Weiner} have provided insight into the phase dynamics of the initial formation of parametric combs, no analysis exists for the phase dynamics of the soliton modelocking process. Although Kerr-based parametric frequency combs have been suggested as the most fundamental example of self-organization in nonlinear optics \cite{Firth, Leo}, no direct connection has been made to the concepts of synchronization and self-organization. \\
\indent In this Letter, we find that phase equations derived from the Lugiato-Lefever equation (LLE) display self-organization features akin to the Kuramoto model, including the existence of meaningful order parameters and coherence-coupling feedback that describes the soliton formation process. These equations predict that the pump phase is offset from the rest of the modal phase profile in the soliton state. Additionally our analysis predicts that phase anti-symmetrization, where the phase profile becomes anti-symmetric about the pump phase, occurs before phase synchronization and soliton formation can occur. We compare the evolution of these equations to that of the full LLE system and observe a strong correspondence between them, indicating that our model captures the dynamics leading to the soliton formation process and explains the role of chaotic states in soliton formation. \\
\indent The governing equation of modelocked parametric frequency combs is the LLE (a damped, driven nonlinear Schr{\"o}dinger equation inside a cavity) with periodic boundary conditions \cite{Coen}:  
\begin{equation}
T_{r}\frac{\partial A}{\partial t}=A_{in}-\big[\frac{\alpha}{2}+i\delta_{o}\big]A+iL\Big[\sum^{3}_{k\geq2}\frac{\beta_{k}}{k!}\left(i\frac{\partial}{\partial \tau}\right)^k+\gamma|A|^2\Big]A. \label{eq:LLE}
\end{equation} 
Here $A$ is the intra-cavity field, $t$ and $\tau$ are the slow and fast times of the system, respectively, $A_{in}$ is the pump field coupled into the cavity at frequency $\omega_0+\delta_{o}$ where $\delta_{o}$ is the detuning of the pump field from the center of the cavity resonance, $\beta_{k}$ are dispersion coefficients, $\gamma$ is the nonlinear coefficient, $\alpha$ represents the total linear loss per round trip of the cavity of length $L$, and $T_{r}=L/v_g$ is the round trip time. We consider the intra-cavity field as a sum of the discrete cavity modes and define a phase $\phi_{p}(t)$ for each mode at the frequency corresponding to the equidistant comb defined by the detuned pump field such that $A(t,\tau)=\sum^{N+1}_{p}A_{p}e^{i(\omega_p+\delta_{o}-\omega_0)t-i(\Omega_{p}-\Omega_{0})\tau}e^{i\phi_{p}(t)}$. By letting the pump mode index $p_0=0$, where N is even and $-N/2\leq p\leq N/2$, the slow and fast frequencies of the field become $\omega_p=2\pi v_g p/L+\omega_0$ and $\Omega_{p}=2\pi v_g p/L+\Omega_0$, where $\omega_0$ and $\Omega_0$ denote the frequencies at the center of the pump resonance and $v_g$ is the corresponding group velocity. For a sufficiently strong pump field $A_{in}(t,\tau)=A_{0}e^{i(\phi_{0}+\delta_{o}t)}$ with appropriate cavity detuning $\delta_{o}$ and constant phase $\phi_{0}$, a broadband comb of frequencies is generated near the modes of the cavity (Fig.\ref{fig:one}c). After an initial build-up period, the amplitudes of the cavity modes reach a near constant state, thus we are able to neglect the amplitude variations of the modes and consider only the phases. By employing a slowly varying envelope approximation for the total intra-cavity field and normalized dispersion coefficients $\xi_{k}=(2\pi v_{g}/L)^{k}v_g\beta_k$, we derive the following general dynamical phase equation from the LLE, with time dependency of the phases made implicit:
\begin{equation}
\dot{\phi_{p}}=\frac{\xi_{2}}{2}p^{2}+\frac{\xi_{3}}{3}p^{3}-\Gamma\sum^{N/2}_{l,m,n = -N/2}\mathscr{A}^{ln}_{mp}\cos(\phi_{l}-\phi_{m}+\phi_{n}-\phi_{p}), \label{eq:LLE_phase}
\end{equation} 
where $\Gamma = \gamma L/T_{r}$ and $\mathscr{A}^{ln}_{mp}=A_{l}A_{m}A_{n}/A_{p}$. This equation has functional similarities to the Kuramoto model wherein each optical mode can be considered an individual oscillator. The spread in natural frequencies of the oscillators is represented by the second- and third-order dispersion terms in the right-hand side, while the nonlinear term gives rise to the coupling among oscillators in the last term in the right-hand side.\\
\indent In the absence of a strong pump mode, this equation has no stable solutions. The cosine coupling term has an equilibrium point at $\pi/2$, which results in the phase mismatch for various FWM processes being pulled to this value. However, this condition cannot be simultaneously satisfied for all combinations of modes. For a pump $\eta$ times stronger than an average comb (non-pump) mode, the coupling term can be decomposed, via the amplitude factor $\mathscr{A}^{ln}_{mp}$, into 4 categories of processes with relative strengths $\eta^2$, $\eta$, 1, $\eta^{-1}$. In our analysis we keep only the largest coupling terms that scale as $\eta^2$ and $\eta$. The terms that scale as $\eta^2$ are a result of the pump-degenerate (PD) FWM processes, where two pump photons are annihilated to create a photon pair at modes symmetric about the pump mode. Alternatively, the terms that scale as $\eta$ are due to the pump-nondegenerate (PND) FWM processes, in which one pump photon and one comb photon are annihilated and create two photons at the energetically appropriate modes. Under this approximation we find that the most natural variables to describe the system are not the individual phases $\phi_{p}$ of the modes, but rather the phase average and difference for pairs of modes symmetric about the pump mode, that is $\bar{\phi}_{p}=(\phi_{p}+\phi_{-p})/2$ and $\theta_{p}=(\phi_{p}-\phi_{-p})/2p$, respectively. We transform to this basis and obtain the following pair of equations, which we term the parametric synchronization equations (PSE): \\
\onecolumngrid
\noindent\rule{18cm}{0.4pt}
\begin{eqnarray}
\dot{\bar{\phi}}_{p} &=& \frac{\xi_2}{2}p^2-2\Gamma\eta^{2}A_{c}^{2}\cos[2(\phi_{0}-\bar{\phi}_{p})]- \Gamma\eta A_{c}^{2}NR(t)\cos(\phi_{0}-\bar{\phi}_{p})\cos[p(\theta_{p}-\theta_{o})], \label{eq:phaseAvg} \\&\nonumber \\
\dot{\theta}_{p} &=&\frac{\xi_3}{3}p^2- \frac{2\Gamma\eta A_{c}^{2}N}{p}R(t)\sin(\phi_{0}-\bar{\phi}_{p})\sin[p(\theta_{p}-\theta_{o})],\label{eq:phaseDiff}
\end{eqnarray}
\noindent\rule{18cm}{0.4pt}
\twocolumngrid
\noindent where $A_{c}=A_{0}/\eta$ is the amplitude of the comb modes, $\phi_{0}$ is the pump phase which is fixed and $N$ is the total number of comb modes in the system. In this transformed basis, the symmetric character of the system is separated from the anti-symmetric character. The phase-average equation describes the symmetric behavior of the system while the phase-difference equation describes its anti-symmetric behavior.  The pump-degenerate processes manifest themselves only in the phase-average equation as the second term on the right-hand side of Eq.(\ref{eq:phaseAvg}). The pump-nondegenerate processes have both symmetric and anti-symmetric contributions and appear as the last terms in the phase-average and phase-difference equations. $R(t)$ and $\theta_{o}(t)$ represent the order parameters of the systems and are given by $R(t)=\frac{2}{N}|\sum^{N/2}_{m=1}e^{im(\theta_{m}-\theta_{o})}|$ and $\theta_{o}(t)=\frac{8}{N^{2}}\sum^{N/2}_{m=1}m\theta_{m}$.
Here $\theta_{o}$ is the normalized average phase difference. It serves the same role as the average phase $\psi$ in the Kuramoto model. It also measures the linear slope of the phase profile which yields a translation of the temporal pulse profile along the cavity length. $R(t)$ is the coherence; it measures the extent to which the population of the phase differences, $\theta_{m}$, aligns to the average phase difference $\theta_{o}$. The triple sum reduces to a single sum since the PND term is only a single sum and the phase-average and phase-difference parameters are separable due to the phase symmetry induced by the PD term. \\
\begin{figure}[t]
\centerline{\includegraphics[width=8.5cm]{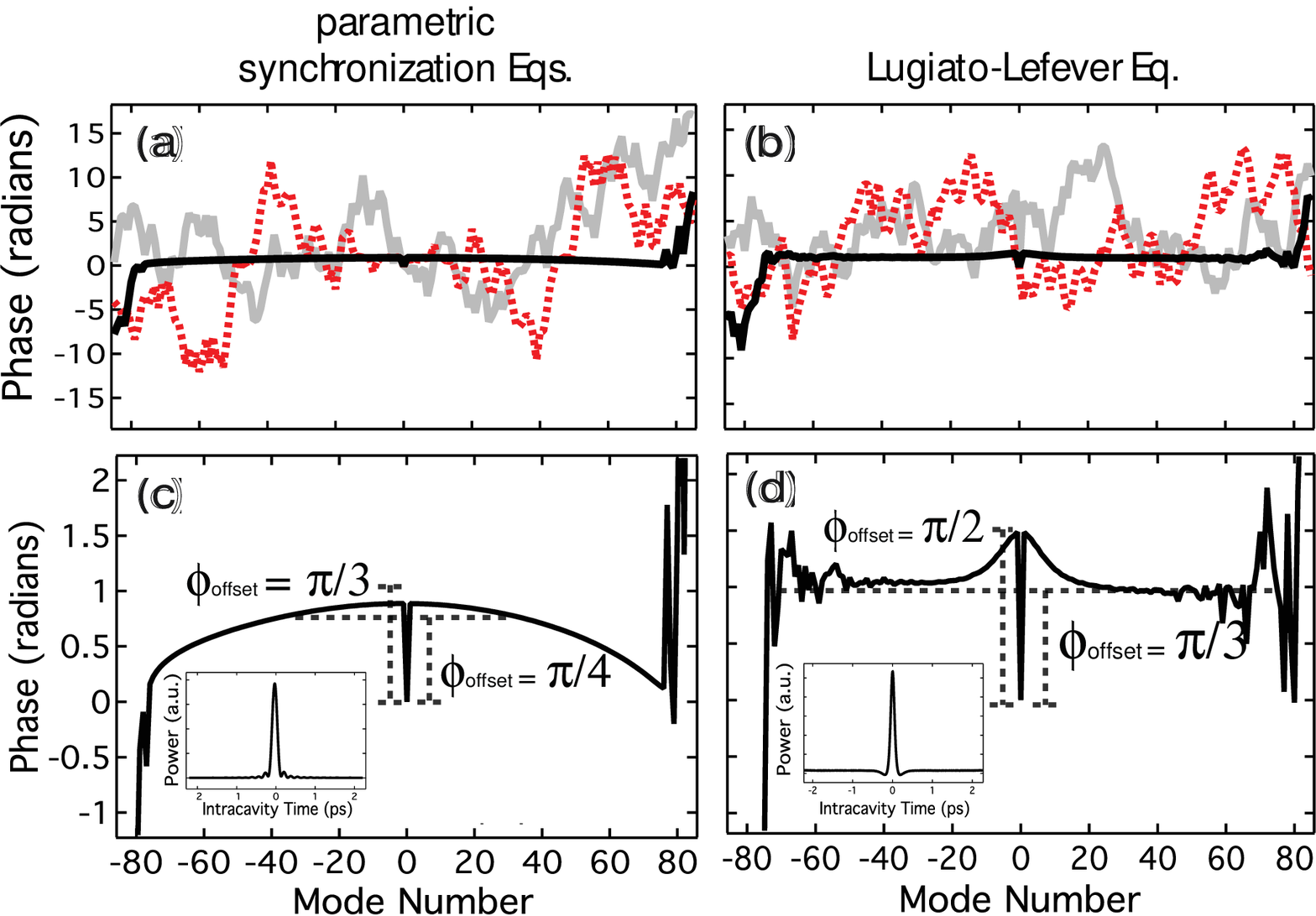}}
\caption{\label{fig:two} (a,b) Three stages of evolution of the phase profile of the intra-cavity field as predicted by (a) the parametric synchronization equations (PSE) and (b) the Lugiato-Lefever equation (LLE): The grey curves represent the initial random phase profile. The red-dotted curves show the phase evolution after 370 (PSE), 308 (LLE) round trips; the phase profiles in both models illustrate the anti-symmetrization of the spectral phase due to the pump degenerate FWM processes. The black curves show the final phase profile after 3394 (both) round trips in which the phases have become completely synchronized, which is a result of the pump non-degenerate terms.  In addition, a slight offset of the phase of the pump from the phases of the other cavity modes is observed. (c,d) The final spectral phase profiles of the PSE and LLE systems showing deviations from a pure linear profile, including the pump phase offsets. (insets) Temporal pulse shapes of the PSE and LLE.}
\end{figure}
\indent We consider the evolution of the PSE system by introducing an initially random phase profile into the equations. Since the PD term scales as $\eta^{2}$ it initially dominates the dynamics, and its presence in the phase-average equation has the effect of anti-symmetrizing the phase profile about the pump phase $\phi_{0}$. The PND term does not initially play a role since the coherence $R(t)$ is zero due to the initially random phases and since it is inherently $\eta/N$ times smaller than the PD terms. Eventually anti-symmetrization of the phases occurs and the coherence becomes non-zero, which allows the PND terms to become non-negligible. This has the effect of synchronizing all the normalized phase differences to their average. This results in a near-linear spectral phase profile, which is consistent with evolution to a cavity soliton as predicted by the LLE model and illustrates its connection to self-organization behavior. Thus, the PD term entrains the phase averages to a fixed input phase, and the PND term employs the coherence-coupling feedback to self-organize around a non-fixed normalized average phase difference. We numerically model the temporal evolution of the PSE and the full LLE systems and verify this prediction through the selected phase profiles in Fig. \ref{fig:two}(a,b). Both models show the progression from initially random phase profiles to an anti-symmetric profile and finally to a fully synchronized profile.\\
\indent One of the key predictions of the PSE is that the system evolves into a state in which the pump phase is offset from the rest of the phase profile. This offset arises from the cosinusoidal dependence of the PD term; in order for this term to act as a restoring force on the phase average as in the Kuramoto model, it must have a sinusoidal dependence on the phase averages. We compare in detail the synchronized spectral phase profiles in Fig. \ref{fig:two}(c,d). Both systems stabilize to a broadband phase-locked state with an offset of the pump phase from the rest of the phase profile. Due to the factor of 2 in the argument of the PD term, this offset should be $0<\phi_{offset}<\pi/2$ and centered at $\pi/4$ in order for the cosine to have a significant sine-like contribution. Both the PSE and the LLE predict pump phase offsets within these bounds. The LLE system has a slightly larger pump phase offset due to self-phase and pump-induced cross-phase modulation effects which were not accounted for in the PSE. This is, to our knowledge, the first theoretical prediction and explanation of the origin of the pump phase offset of the soliton-modelocked states in a parametric frequency comb. We confirm in Fig. \ref{fig:two}(c,d insets) that the broadband phase-locked state results in a solitary pulse in the time domain. The exact pulse shape for the PSE is not quantitatively meaningful since all the modes have equal amplitude resulting in a Sinc-like pulse without a CW background.\\
\indent Figure \ref{fig:four}(a,b) compares the evolution of the order parameters, the coherence, and the normalized average phase difference in the PSE and LLE systems. Despite slight quantitative differences, both systems exhibit abrupt transitions from a disordered state to an ordered state as indicated by the sharp rise and subsequent stabilization of the coherence. The normalized average phase difference also exhibits dynamical similarities in the initial rapid increase and subsequent decline and stabilization to a constant value. Closer inspection of the two parameters (insets) reveal relaxation behavior on the order of 1 ns, close to the cavity lifetime of 1.42 ns. It's also insightful to consider the phase symmetry of the two systems as quantified by $R_{sym}= \frac{2}{N}|\sum^{N/2}_{m=1}e^{i(\bar{\phi}_{m}-\phi_{0})}|$. This value is a measure of the extent to which the phase profile is anti-symmetric about the pump phase and is equivalent to a coherence of the phase averages. Figure \ref{fig:four}(c,d) compares the coherence and the phase symmetry for the two systems. In both systems phase symmetry occurs before coherence is achieved, and the coherence cannot grow until the phase symmetry has reached a high value. This confirms our initial prediction that the PD term must anti-symmetrize the phase profile before the PND term can synchronize the phases to a near linear profile. Furthermore, the phase symmetry does not fully stabilize until the coherence has reached a high value, and in turn, the coherence does not stabilize until the phase symmetry has fully stabilized. These results illustrate the necessity of phase anti-symmetrization to precede phase synchronization in soliton formation and the complex interplay between phase symmetry and phase coherence.\\
\indent Lastly, the phase evolution modeled in the LLE is the stage after it evolves from the chaotic state and enters a soliton state. Passage through a chaotic state has been suggested as necessary for soliton modelocking to occur \cite{Lamont, Coillet2}. The correspondence between Fig. \ref{fig:four}(a,c) and (b,d) indicates that the PSE captures the essence of the soliton formation process in parametric frequency combs. It also supports the conclusion that the chaotic stage is necessary in soliton formation for the purpose of randomizing the phase profile to prevent Turing pattern and mini-comb-related FWM processes from dominating the phase matching of the comb \cite{Coillet, Herr}. Turing states and the associated mini-combs have phase profiles that inherently lack global symmetry about the pump phase and thus cannot directly enter into a soliton state. However, some parameter regimes of the PSEs have produced phase profiles with multiple phase-offset modes and phase-steps, similar to those states that have been measured by Del'Haye \emph{et al.} \cite{DelHayeHot}, and are the subject of ongoing work. \\
\indent In saturable absorption based modelocked lasers the cosine in Eq.\ref{eq:LLE_phase} is replaced by a sine and phase synchronization is possible without a strong coherent pump field. Given the generality of the root equation of the LLE, which is the complex Ginzburg-Landau equation, this synchronization model may be applicable to the phase transition dynamics in a wide range of physical systems.\\
\indent The authors would like to acknowledge Y. Okawachi, R. Lau and M. Yu for insightful discussions and the Defense Advanced Research Projects Agency (DARPA) and Air Force Office of Scientific Research (AFOSR) for supporting this work.



\end{document}